\documentclass{aa}
\usepackage{amsmath}
\usepackage{txfonts}
\usepackage{times}
\usepackage{graphicx}
\usepackage{natbib}

\begin{document}

\def\bhm{M_{\text{BH}}} 
\def\dotm{\dot{m}}
\def\dotM{\dot{M}}
\def\dotMedd{\dot{M}_{\rm Edd}}
\def\dv{\Delta v}
\def\integral{{\it{INTEGRAL}~}}
\def\Gj{\Gamma_{10}}
\def\ledd{L_{\rm Edd}}
\def\luvext{L_{\rm UV, 45}^{\rm ext}}
\def\lxext{L_{\rm X, 45}^{\rm ext}}
\def\mbh{m_{\rm bh}}
\def\ooo{[O\,{\sc iii}]}
\def\nth{n_{\rm th}}
\def\st{\sigma_{\rm T}}
\def\ud{U_{\rm D}}
\def\udp{U_{\rm D}^{\prime}}
\def\usyn{U_{\rm syn}^{\prime}}
\def\uth{U_{\rm th}}
\def\xiluv{\left(\xi L_{\rm UV}\right)_{45}}

\def\fexxv{Fe {\sc xxv}} 
\def\fexxvi{Fe {\sc xxvi}}
\def\nex{Ne {\sc x}}
\def\sixiv{Si {\sc xiv}} 

\title{A Possible Feature of Thermal Matter in Relativistic Jets of Radio-loud
Quasars}

\author{J.-M. Wang\inst{1,2,3}, 
        R. Staubert\inst{1}
        and T.J.-L. Courvoisier\inst{4,5}}

\authorrunning{J.-M. Wang, R. Staubert \& T.J.-L. Courvoisier}
\titlerunning{Thermal Matter in Relativistic Jets of Radio-loud Quasars}
\offprints{J.-M. Wang\\ \email{wang@astro.uni-tuebingen.de}}

\institute{Institut f\"ur Astronomie und Astrophysik, Abt. Astronomie,
Universit\"at T\"ubingen, Sand 1, D-72076 T\"ubingen, Germany,
\and Laboratory for High Energy Astrophysics, Institute of High
Energy Physics, CAS, Beijing 100039, P. R. China.
\and Alexander von Humboldt Fellow
\and INTEGRAL Science Data Center, Chemin d'Ecogia 16, CH-1290 Versoix, Switzerland.
\and Geneva Observatory, 51 ch. des Maillettes, CH-1290 Sauverny, Switzerland.}

\date{Received 11 November 2003 / Accepted 6 April 2004}

\abstract{It has been suggested that relativistic jets in quasars may 
contain a considerable amount of thermal matter. In this paper, 
we explore the possibility
that the K$\alpha$ line from the thermal matter may appear at tens of
keV due to a high Doppler blue-shift. 
In the jet comoving frame,
the energy density of photons originally emitted by the accretion disk
and reflected off the broad line region clouds dominates over that
of photons of other origin.
We discuss the photoionization states of the thermal matter and find 
that the irons elements are neutral.
The high metallicity in quasars enhances the possibility to detect 
the thermal matter in the relativistic 
jet in some radio-loud quasars. A highly Doppler blue-shifted K$\alpha$
line may be detected. We make a prediction for 3C~273, 
in which the K$\alpha$ line luminosity might be of the order 
$3.0\times 10^{44} {\rm erg~s^{-1}}$ with an equivalent width of 2.4~keV. 
Such a line could be detected in a future mission. 
%
}

\maketitle

\section{Introduction}

It is well known that the observed powerful relativistic jets are an 
important ingredient in radio-loud quasars. 
The non-thermal electrons are responsible for the
multiwaveband radiation although the specific acceleration mechanism
remains a matter of debate. Little 
attention has been paid to questions related to the thermal matter in 
the relativistic jets.

As an example of a galactic jet source,
SS433 has been extensively studied both observationally and 
theoretically (Kotani et al. 1997; Brinkmann \& Kawai 2000). {\it Chandra} 
discovered several lines from highly ionized atoms, such as \ion{Fe}{xxv}, 
Fe {\sc xxiv}, Co {\sc xiv}, S {\sc xvi}, Ly$\alpha$ and Ly$\beta$, 
Ne {\sc x} and Mg {\sc xi} etc. (Marshall et al. 2002). The chemical 
composition in these jets definitely includes heavy elements rather than 
pure pair plasma. This could be explained by a model in which the emission
lines are originally from the hot plasma expanding in the jet 
(Brinkmann \& Kawai 2000; Memola et al. 2002).   
In radio-loud quasars, the case is highly uncertain. A heavy jet 
mainly composed of proton-electron plasma 
has been suggested by Celotti \& Fabian (1993) based on the kinetic 
luminosities of jets found in a large VLBI sample. The absences of
soft X-ray bumps in radio-loud quasars lead to the exclusions of a pure 
pair and pure proton-electron plasma, most likely, the relativistic jet is
pair-dominated numberwise but still dynamically dominated by 
protons (Sikora \& Madejski 2000). The measurement of polarization in
a few objects seem to favor the pair plasma (Wardle et al. 1999,
Hirotani et al. 1999), but the linear polarization strongly supports a 
normal plasma as the main composition (Fraix-Burnet 2002).
Ruszkowski \& Begelman (2002) find that the electron-proton
and electron-positron jets can lead to the same circular and linear 
polarization in 3C 279.

There is growing interest in the presence of thermal matter in 
relativistic jets (Celotti et al. 1998). The thermal matter, as
argued by Celotti et al. (1998), may be due to: 1) not 100\%
matter can be accelerated to relativistic energy; 2) non-thermal
matter cools down and is thermalized before being reaccelerated; 3) some 
thermal matter might also be trapped at the base of the jet as they form 
and some are loaded by the surrounding external medium (but see
Lyutikov \& Blandford 2002 for a different view). Kuncic et al.
(1997) used {\sc cloudy} to make detailed calculations 
of emission lines from the thermal matter immersed in the non-thermal 
radiation field in the jet. The basic features are the presence of 
emission lines in the extreme ultraviolet band. However, the situation
in radio-loud quasars may not be so simple. 
There are three types  of possible sources for the {\it ionizing} 
photons in the 
relativistic jet: 1) synchrotron photons (Blandford \& K\"onigl 1979); 
2) accretion disk photons (Melia \& K\"onigl 1989; Dermer \& 
Schlickeiser 1993); 3) diffuse photons in the broad line region (Sikora
et al. 1994).  At present the observations do not allow 
to decide which might dominate in blazars. 
In the jets comoving frame the thermal matter sees photons
originating from the disk and reflected by the
clouds in the broad line region. These photons may dominate over the local 
synchrotron photons (Sikora et al. 1994). Moreover, it is
of great interest to note that a high metallicity is common in quasars 
(Hamann \& Ferland 1999).

In this paper, we show that the thermal matter in
the jet is neutral and the observational features 
of the thermal matter will mainly be the presence of a highly Doppler 
shifted K$\alpha$ line, which could be detected in some
radio-loud quasars by future instruments. 

\section{Emission from Thermal Matter in Jets}
\subsection{Ionization of thermal matter and Fe K$\alpha$ line}

Detailed constraints on temperature, density and size of the 
thermal matter in relativistic jets have been discussed by Celotti 
et al. (1998), who suggest that the thermal matter may exist as cold
clouds. The microscopic properties of the surviving clouds in 
the relativistic jet depend on many processes. The density and size
of the clouds are largely uncertain. In this paper, we take
typical values for the density, temperature and the size as given by 
Celotti et al. (1998), e.g. $n_t=10^{14}n_{\rm t, 14}$ cm$^{-3}$,
$T=10^5T_5$K and $R_{\rm th}=10^9R_{\rm th, 9}$~cm. 
For illustration, we consider the clouds in the jet at a distance $R$
from the central black hole, with $R=10R_g=1.5\times 10^{15}$cm, where 
$R_g=1.5\times 10^{14}({\rm cm}) m_9$ and $m_9=M_{\text{BH}}/10^9M_{\odot}$. 
We approximate the Doppler factor ${\cal D}$ with the Lorentz factor 
$\Gamma$ and take $\Gj=\Gamma/10\approx 1$. 

The peak frequency of the thermal emission from such clouds 
in the jet will be
$\Gamma kT\approx 0.1~({\rm keV}) \Gj T_5$ in the observer's frame, 
where $k$ is Boltzmann constant. If such a soft X-ray emission is absent 
in the observed continuum, the number of the cold clouds $N_0$ should
be constrained by  
\begin{equation}
N_0\le 5.62\times 10^8 L_{\rm SX, 46}\Gj^{-4}T_5^{-4}R_{\rm th,9}^{-2}.
\end{equation}
where the soft X-ray luminosity $L_{\rm SX}=10^{46} L_{\rm SX, 46}~{\rm erg~s^{-1}}$.
Assuming the jet geometry to be a cone with opening angle 
$\theta_j\sim 1/\Gamma$, the cross section radius of the jet is 
$a=\theta_j R$ . 
Since the sound velocity inside the cold clouds is much less than that in the
intercloud medium, the expansion of the clouds can be neglected.
The filling factor of the thermal clouds in the jet volume is 
$f\approx \sum_i^{N_0}\left(4\pi/3\right)R_{\rm th}^3/\left(\pi R^3/3\Gamma^2\right)
\approx 6.66\times 10^{-8}~ \Gj^{2}R_{\rm th,9}^3R_{15}^{-3}$,
where $R_{15}=R/10^{15}$cm and
the opening angle is taken as $\theta\approx 1/\Gamma$, indicating
that the jet is very clumped with thermal clouds as suggested 
by Celotti et al. (1998). These general constraints on the thermal 
matter are in agreement with those of Sikora et al. (1997).

According to the disk-corona model by Haardt \& 
Maraschi (1993), most of the gravitational energy will be released in the
hot corona as X-ray emission, and UV emission as reprocessed 
X-rays. We take a similar scenario for the accretion disks
in flat spectrum radio quasars. As an example, we take the accretion rate 
$\dot{m}=L_{\rm bol}/L_{\rm Edd}=0.2$, where the Eddington luminosity is
$L_{\rm Edd}=1.26\times 10^{47}m_9{\rm (erg~s^{-1})}$. The bolometric luminosity 
is then $L_{\rm bol}=2.52\times 10^{46}{\rm (erg~s^{-1})}$.
For simplicity we assume that
half of the bolometric luminosity is released in X-rays, 
$L_{\rm X}=1.26\times 10^{46}m_9{\rm (erg~s^{-1})}$, and the other half in 
optical/UV, $L_{\rm UV}=1.26\times 10^{46}m_9{\rm (ergs~s^{-1})}$. Some of 
the disk emission will be reflected by the clouds in the broad line 
region. 

Sikora et al. (1994) point out that the photons in the relativistic jet received 
from diffuse scattering of accretion disk photons in the BLR may dominate over 
the local synchrotron photons and the photons received directly from the disk. 
The nature of the BLR clouds remains open (Alexander \& Netzer 1994; Baldwin et al. 
2003), the covering factor $\xi\approx 0.1 $ is indicated by the energy budget of 
emission lines from BLR clouds (Netzer 1990).
The typical density and  temperature are $n_e=10^{11\sim 12}$cm$^{-3}$ and 
$T=10^{4\sim 5}$K in
the clouds of BLR, and the ionized fraction of dimension is typically $\ell=10^{12}$cm. 
The Thomson scattering optical depth is
$\tau_{\rm es}\equiv n_e\st\ell \sim 0.1$, where $\st$ is the Thomson cross section.
Collin-Souffrin et al. (1996) calculate emergent spectra from an optically thin cloud 
(with solar abundances) radiatively heated in detail. They found that 
about 1\% of the incident radiation 
will be reflected for a cloud with column density $\Sigma=10^{22}$g~cm$^{-2}
($Thomson depth $\tau_{\rm es}\approx 0.006$) and the reflected amount is insensitive 
to the cloud's temperature. In the optically thin regime, the reflected flux will be 
approximately proportional to the optical depth. Therefore, the reflected fraction of
the incident radiation from disk is ${\cal R}> 0.1$ for the BLR clouds with
$\tau_{\rm es}\approx 0.1$.
The energy density of the diffuse X-rays is $U_{\rm X}=L_{\rm X}^{\rm ref}/4\pi R_{\rm BLR}^2c$,
where $L^{\rm ref}_{\rm X}=\xi {\cal R} L_{\rm X}$, $R_{\rm BLR}$ is the scale of the BLR
and $c$ the light speed.
In the jet frame, the energy density of the received X-rays by the thermal matter is then
\begin{equation}
U_{\rm X}^{\prime}=\Gamma^2U_{\rm X}
                  =0.9~({\rm erg~cm^{-3}})\Gj^2L^{\rm ref}_{\rm X,44}
                   R_{\rm 0.1pc}^{-2},
\end{equation}
where we take the reflected fraction of hard X-rays $\xi {\cal R}=0.01$ conservatively, 
the BLR size $R_{\rm 0.1pc}=R_{\rm BLR}/$0.1pc and 
$L_{\rm X,44}^{\rm ref}=L_{\rm X}^{\rm ref}/10^{44}{\rm erg~s^{-1}}$.
The ionization parameter defined by 
\begin{equation}
\Xi=\frac{U_{\rm X}^{\prime}}{n_{\rm t} k T}
\approx 6.0\times 10^{-4}~\Gj^2L_{\rm X,44}^{\rm ref}R_{\rm 0.1pc}^{-2}n_{\rm t, 14}^{-1}T_5^{-1},
\end{equation} 
indicating that most of the atoms are neutral (Krolik \& Kallman 1984)
and the iron K$\alpha$ line is at 6.4keV. It is thus expected that 
the thermal clouds will maintain their temperatures.

\subsection{Iron abundance and line luminosity}
As shown by Monte Carlo simulation of Reynolds (1996), the iron K$\alpha$
line is the most prominent among the emergent lines in the reprocessed spectrum. 
We neglect the lines of other elements since they are much fainter
than the iron K$\alpha$ line.  We take the relative abundance of iron as
${\cal A}=Z_{\rm QSO}/Z_{\odot}$, where $Z_{\rm QSO}$ is
the iron abundance of QSO and $Z_{\odot}=3\times 10^{-5}$. 
Krolik \& Kallman (1987) (see also Liedahl 1999 for details) show
that the Fe K-edge opacity is
\begin{equation}
\kappa_{\rm K_0}= 0.67\kappa_{\rm es}f(\Xi){\cal A}
          \left(\frac{E}{E_{\rm K_0}}\right)^{-3}
\end{equation}
where $\kappa_{\rm es}=0.34$ is the Thomson scattering opacity, and 
$f(\Xi)$ is a slow function of $\Xi$. The optical depth of one cold 
cloud is
\begin{equation}
\tau_{\rm K}=0.45f(\Xi){\cal A}_{10}n_{\rm t, 14}R_{\rm t,9},
\end{equation}
where ${\cal A}_{10}={\cal A}/10$.  For simplicity we assume that the 
reflected spectrum as the photoionization source is in a simple form as 
$L_{E}^{\prime}=L_0^{\prime}\left(E/E_0\right)^{-\gamma}$
within $E_0\le E \le E_1$,
where $\gamma\approx 0.7$ is the index of the reflected spectrum by 
the clouds in the BLR. In the jet's comoving frame, 
$\int L_{E}^{\prime}dE=\Gamma^2L^{\rm ref}_{\rm X}$, 
yielding
$L_0^{\prime}=(1-\gamma)C_0E_0^{-1}\Gamma^2L^{\rm ref}_{\rm X}
\left(E_1/E_0\right)^{-\gamma}$, and 
$C_0=\left[\left(E_1/E_0\right)^{1-\gamma}-1\right]^{-1}$. 
The reflected spectrum is assumed to be the same as the incident radiation
field.  This may be not exact, but the accuracy is enough for the present goal
since the K$\alpha$ line flux mainly depends on the total energy of hard X-rays.

With the optical depth given by equation (5), 
the luminosity of the K$\alpha$ line from $N_0$ clouds 
in the jets comoving frame will be given by
\begin{equation}
L_{\rm K\alpha}^{\prime} \approx \langle Y\rangle \tau_{\rm K}L_0^{\prime}
\frac{E_{\rm K\alpha}}{\gamma+3}\left(\frac{N_0\delta \omega}{4\pi}\right)
\end{equation}
where $\langle Y\rangle$ is the fluorescent yield for the production of 
K$\alpha$, $\delta \omega$ is the solid angle of one cloud subtended 
at the X-ray continuum source. In the observer's frame, the received 
luminosity is $L_{\rm K\alpha}^{\rm obs}=\Gamma^4L_{\rm K\alpha}^{\prime}$, i.e.
\begin{equation}
L_{\rm K\alpha}^{\rm obs}=3.0\times 10^{44}~({\rm erg/s})\Gj^6{\cal A}_{10}
          n_{\rm t, 14}R_{\rm th, 9}^3R_{15}^{-2}L_{\rm X, 44}^{\rm ref}\\
\end{equation}
with $\langle Y\rangle \approx 0.6$, $f(\Xi)=1.7$, $E_0\approx E_{\rm K\alpha}$,
$E_1/E_0=100/6.4$ and $N_0=5.62\times 10^8$. The observed energy of the K$\alpha$ line 
will be $E_{\rm K\alpha}^{\rm obs}\approx 64~({\rm keV})~\Gj$.

The profile of such a line may be mainly broadened by the relative motion among the cold
clouds. As argued by Celotti et al. (1998), the strength of a comoving
magnetic field is of typical value 
$B=2\times 10^3L_{\rm jet,46}^{1/2}R_{15}^{-1}\Gamma_{10}^{-1}$
Gauss, where $L_{\rm jet,46}=L_{\rm jet}/10^{46}~{\rm erg~s^{-1}}$ is the power of the jet
as Poynting flux. Such a magnetic field can  confine the relative motion among the cold 
clouds with respect 
to the relativistic bulk flow, otherwise the collimation of the jet will be broken down. 
This random velocity of the clouds $\upsilon_c$ can be estimated 
by $\frac{1}{2}n_tm_p\upsilon_{\rm c}^2=B^2/8\pi$. For the typical value, we have 
$\upsilon_c/c=1.5\times 10^{-3}$, namely,  
$\Delta E\approx \Gamma E_{\rm K\alpha}\upsilon_c/c\approx 0.1\Gj$keV.
If the thermal clouds follow the opening angle of the jet due to random motion, on the 
other hand, their relative velocity would be of order of 
$\upsilon_c/c\approx 0.1\Gamma_{10}^{-1}$. This causes a broadening of
$\Delta E\approx 6.4$keV. The resolution of the line profile may probe more detail
dynamics of the jet in future.

From eq. (7), we see that the observed line luminosity is very 
sensitive to the Lorentz factor $\Gamma$ and proportional to the iron abundance 
${\cal A}$. We use the maximum number of cold clouds (see equation 1), the predicted
luminosity of the iron K$\alpha$ line is the upper limit.  This limit is due to the 
absence of features of bulk Comptonization in the soft X-ray band. We note that the 
above model only works within BLR ($\sim 0.1$pc).

\subsection{Candidate objects with a blue-shifted K$\alpha$ line}
The proposed model naturally relates to the $\gamma$-ray emission 
model advocated by Sikora et al. (1994). 
$L_{\gamma}/L_{\rm syn}\approx U^{\rm ref}_{\rm UV}/U_{\rm B}$ and 
$U^{\rm ref}_{\rm UV}>U_{\rm syn}$ implies that 
$L_{\gamma}/L_{\rm syn}>1$ if the equipartition between magnetic 
field and relativistic electrons $U_{\rm B}=U_{\rm e}$ is fulfilled.
Here $L_{\gamma}$ is the $\gamma$-ray luminosity, $L_{\rm syn}$ is 
the luminosity due to synchrotron emission, $U_{\rm B}$ and 
$U_{\rm UV}^{\rm ref}$ are the energy densities of the magnetic 
field and of the reflected UV luminosity, respectively. 
{\it CGRO} observations of $\gamma$-rays from blazars show 
the importance of reflection (Dondi \& Ghisellini 1995). 
The potential candidates for a blue-shifted iron K$\alpha$ 
line should be those objects in which 
$L_{\gamma}/L_{\rm syn}>1$. On the other hand 
the present model 
needs metal-rich thermal matter. This condition corresponds to the 
observable indicator of metallicity as N {\sc v}/C {\sc iv}$\ge 0.4$ 
(Hamann \& Ferland 1999). We thus have the criteria for potential 
candidates as
\begin{equation}
\frac{L_{\gamma}}{L_{\rm syn}}>1;~~~{\rm \frac{N~V}{C~IV}\ge 0.4}.
\end{equation}
These conditions are likely
satisfied in flat spectrum radio quasars. The superluminal motion of
$\gamma$-ray bright blazars show that $\Gamma\sim 10$ is common
among the $\gamma$-ray loud AGNs (Jorstad et al. 2001). 
In the sample of Dondi \& Ghisellini (1995), the mean ratio of 
$L_{\gamma}/L_{\rm op}\approx 30$. We expect that the highly
Doppler-shifted K$\alpha$ line appears in most of these
objects with high flux ratio N {\sc v}/C {\sc iv} in the sample of
Dondi \& Ghisellini (1995).

\section{A predicted blue-shifted K$\alpha$ line in 3C 273}
3C~273 is a typical
blazar with strong emission lines, powerful continuum emission 
from radio to $\gamma$-rays (see an extensive review of Courvoisier 1998).
A very strong MeV bump has been discovered in 3C~273 by the
{\it Compton Gamma-ray Observatory} (Lichti et al. 1995). This MeV
feature could be explained by several different models, for example,
a break in the electron injection function (Ghisellini et al. 1996),
incomplete cooling of relativistic electrons (Sikora et al. 1994);
and pair cascade process (Blandford \& Levinson 1995). However, in such
models, one encounters other difficulties to reconcile the broad band 
multi-wavelength continuum emission [see a brief review of Sikora et al. (1997)]. 
As argued by Sikora et al. (1997), the MeV feature
could be explained naturally by the ``hot electrons'' version of the
external radiation Compton model (Sikora, Begelman \& Rees 1994)
provided the plasma is not dominated by pairs.
The future detection of the highly blue-shifted K$\alpha$ line could help
clarify the composition of the relativistic jet in radio-loud quasars.

The mass of the black hole in 3C 273 can be estimated from
the absolute magnitude $M_R$ of its host galaxy via 
$\log M_{\rm BH}/M_{\odot}=-0.5M_R-2.96$, 
$M_{\rm BH}=1.2\times 10^9M_{\odot}$ from $M_R=-24.4$ (McLure 
\& Dunlop 2001). The accretion rate of the black hole can be 
roughly estimated from the big blue bump (Walter et al. 1994, 
Wang \& Zhou 1996). We find $\dot{m}\approx 0.4$, about half 
the Eddington luminosity, similar to Courvoisier (1998). The
metallicity can be estimated through the flux ratio of N {\sc v}/C {\sc iv} 
(Hamann \& Ferland 1999). The observed flux ratio of N {\sc v}/\ion{C}{iv}
is 0.46 in 3C 273 (Baldwin et al. 1989), which gives a metal
abundance ${\cal A}=Z/Z_{\odot}\approx 10$ from Fig.~6 of Hamann \& 
Ferland (1999). The superluminal motion has been extensively
studied, the latest observation shows that the apparent velocities
for different components are from 9$c$ to 22$c$ 
(Jorstad et al. 2001). Here, we take $\Gamma\approx 10$.

For the parameters of 3C 273, we estimate a K$\alpha$ line luminosity
$ L_{\rm K\alpha}^{\rm obs}\approx 3.0\times 10^{44}~{\rm erg~s^{-1}}$ and
the flux $F_{\rm K\alpha}=3.9 \times 10^{-12}~{\rm erg~cm^{-2}~s^{-1}}$. 
For the equivalent width, 
we use the {\em INTEGRAL} continuum spectrum, $F_E=3.6\times 10^{-11}E^{-0.73}$
erg~cm$^{-2}$~s$^{-1}$~keV$^{-1}$ (Courvoisier et al. 2003).
The equivalent width is given by
$EW({\rm K\alpha})\approx F_{\rm K\alpha}/F_{E}
                 \approx 2.4~({\rm keV})$.

\begin{figure}
\centerline{\includegraphics[angle=-90,width=8.0cm]{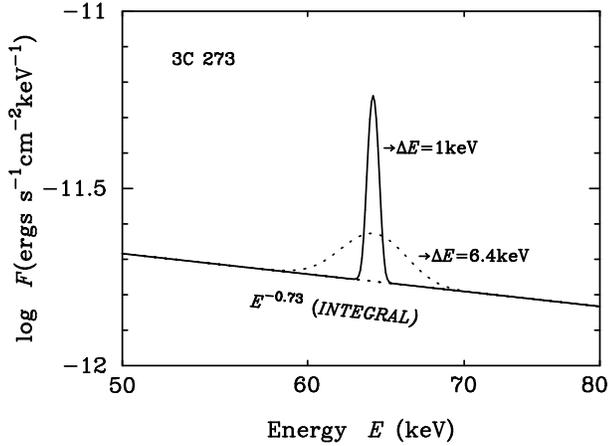}}
\caption{\footnotesize
A plot of the line profiles with future missions.
A Gaussian profile for the line 
is assumed in the jet comoving frame.
We assume two cases with energy resolution of
$\Delta E=1$keV and $\Delta E=6.4$keV, respectively.
The underlying continuum is the {\it INTEGRAL} spectrum 
observed by Courvoisier et al. (2003). 
}
\label{fig1}
\end{figure}


Using the observed background level for {\it SPI}, both analytical
calculations as well as Monte Carlo simulations show that {\it SPI}
will not be able to detect such a line in a total observing time of 
one million seconds. Figure 1 shows that for a next generation instrument for 
which we assume a 1 keV energy resolution a signal to noise ratio of $\sim 1-3$ 
per resolution element is necessary in order to detect the line. This represents 
an improvement by a factor of $3\sim 4$ on the presently
available data that we expect will be available in missions like {\em NeXT}
(New X-ray Telescope, Takahashi et al. 2004) and {\em XEUS} (X-ray Evolving
Universe Spectrometer). The line profile for a well-collimated jet
with the maximum width ($\Delta E=6.4$keV) discussed in \S 2.2 is also plotted in 
Figure 1. 

\section{Conclusions}
We show that future instruments may
allow us to detect a thermal line emitted by matter in a jet directed
towards us and therefore shifted to the blue by a factor that reflects the
gamma factor of the jet. This would allow a direct measurement of the jet
gamma factor and give very important indications on the as of yet not clear
nature of the jet. The line flux is determined by 3 factors: the ionization
level, the Lorentz factor and the metal abundance. Using realistic values
for these parameters, we show that the next generation hard X-ray instruments
may well measure this component.

\acknowledgements{The authors are grateful to the anonymous referee for pointing out
an error in the early version of the paper and 
the helpful comments significantly improving the manuscript. They thank M. 
Stuhlinger for useful
discussions and V. Beckmann for performing SPI simulations. H. Netzer is greatly
thanked for helpful suggestions. J.-M.W. acknowledges support from the Alexander 
von Humboldt Foundation, NSFC and the Special Funds for Major State Basic
Research Project.}


\begin{thebibliography}{99}
\bibitem{}
Alexander, T. \& Netzer, H., 1994, MNRAS 270, 781

\bibitem{}
Baldwin, J.~A., Ferland, G.J., Korista, K.T., Hamann, F. \& Dietrich, M., 
2003, ApJ, 582, 590

\bibitem{}
Baldwin, J.~A., Wampler, E.~J., \& Gaskell, C.~M., 1989, ApJ, 338, 630

\bibitem{}
Blandford, R. \& Levinson, A., 1995, ApJ, 441, 79 

\bibitem{}
Blandford, R. \& K\"ongl, A., 1979, ApJ, 232, 34

\bibitem{}
Brinkmann, W. \& Kawai, N., 2000, A\&A, 363, 640


\bibitem{}
Celotti, A. \& Fabian, A.~C., 1993, MNRAS, 264, 228

\bibitem{}
Celotti, A., Kuncic, Z., Rees, M.~ J. \& Wardle, J.~F.~C,
1998, MNRAS, 293, 288

\bibitem{}
Collin-Soufrrin, S., Czerny, B., Dumont, A.-M \& \.Zycki, P.T., 1996, A\&A, 314, 393

\bibitem{}
Courvoisier, T.~J.~-L., 1998, A\&ARv, 9, 1

\bibitem{}
Courvoisier, T.~J.~-L.; Beckmann, V.; Bourban, G. et al.,
2003, A\&A, 411, L343 

\bibitem{}
Dermer, C.D., \& Schlickeiser, R., 1993, ApJ, 416, 458

\bibitem{}
Dondi, L. \& Ghisellini, G., 1995, MNRAS, 273, 583

\bibitem{}
Fraix-Burnet, D., 2002, A\&A, 381, 374

\bibitem{}
Ghisellini, G., Maraschi, L. \& Dondi, L., 1996, A\&AS, 120, 503

\bibitem{}
Hamann, F. \& Ferland, G., 1999, ARAA, 37, 487

\bibitem{}
Haardt, F. \& Maraschi, L., 1993, ApJ, 413, 507

\bibitem{}
Hirotani, K., Iguchi, S., Kimura, M.,  Wajima, K., 1999, PASJ, 51, 263

\bibitem{}
Jorstad, S.~G., Marscher, A.~P., Mattox, J.~R., Wehrle, A.~E., Bloom, S.~D., 
Yurchenko, A.~V., 2001, ApJS, 134, 181


\bibitem{}
Kotani, T.,  Kawai, N., Matsuoka, M., \& Brinkmann, W., 1996, PASJ, 48, 619

\bibitem{}
Krolik, J.H. \& Kallman, T.R., 1984, ApJ, 286, 366 

\bibitem{}
Krolik, J.H. \& Kallman, T.R., 1987, ApJ, 320, L5

\bibitem{}
Kuncic, Z., Celotti, A. \& Rees, M.~J., 1997, MNRAS, 284, 717

\bibitem{}
Lichti, G. G.; Balonek, T.; Courvoisier, T. J.-L. et al.,
1995, A\&A, 298, 711


\bibitem{}
Liedahl, D.A., 1997, X-ray Spectroscopy in Astrophysics, ed. J. Paradijis \&
     J.~A.~M Bleeker, Springer, 189

\bibitem{}
Lyutikov, M. \& Blandford, R., 2002, astro-ph/0210671


\bibitem{}
Marshall, H.~L., Canizares, C.~R., Schulz, N.~S., 2002, ApJ, 564, 941

\bibitem{}
McLure, R.J., \& Dunlop, J.S., 2001, MNRAS, 327, 199

\bibitem{}
Melia, F.  \& K\"onigl, A., 1989, ApJ, 340, 162

\bibitem{}
Memola, E., Fendt, Ch. \& Brinkmann, W., 2002, A\&A, 385, 1089

\bibitem{}
Netzer, H., 1990, in Active Galactic Nuclei, ed. R. Blandford, H. Netzer \&
  L. Woltjer (Berlin: Springer), 57


\bibitem{}
Reynolds, C.~S., 1996, PhD thesis, University of Cambridge.

\bibitem{}
Ruszkowski, M. \& Begelman, M.C., 2002, ApJ, 573, 485


\bibitem{}
Sikora, M., Begelman, M.C. \& Rees, M.J., 1994, ApJ, 421, 153

\bibitem{}
Sikora, M. \& Madejski, G., 2000, ApJ, 534, 109

\bibitem{}
Sikora, M., Madejski, G., Moderski, R. \& Poutanen, J., 1997, ApJ, 484, 108

\bibitem{}
Takahashi, T., Makishima, K., Fukazawa, Y., et al., 
2004, New Astr., 48, 269 

\bibitem{}
Wardle, J.~F.~C., Homan, D.~C.; Ojha, R.; Roberts, D.~ H., 1998, Nat, 395, 457

\bibitem{}
Walter, R.,  
Orr, A., Courvoisier, T. J.~-L., Fink, H.~H., Makino, F., Otani, C., \& Wamsteker, W., 
1994, A\&A, 285, 119

\bibitem{}
Wang, J.-M., \& Zhou, Y.-Y., 1996, ApJ, 469, 564 


\end{thebibliography}
\end{document}